\documentclass[superscriptaddress,onecolumn,showpacs,showkeys,amsmath,preprint]{revtex4}

\usepackage{graphicx}
\usepackage{dcolumn}
\usepackage[sort&compress]{natbib}
\usepackage{subfigure}
\usepackage{hyperref}

\begin{document}
\title{Effects of boundaries in mesoscopic superconductors}
\author{Antonio R. de C. Romaguera}%
\affiliation{Universiteit Antwerpen, Groenenborgerlaan 171, B-2020 Antwerpen, Belgium}%
\affiliation{Universidade Federal do Rio de Janeiro, Caixa Postal:68.528, Rio de Janeiro - RJ, 21941-972, Brazil}%
\author{Mauro M. Doria}%
\email{mmd@if.ufrj.br}%
\affiliation{Universiteit Antwerpen, Groenenborgerlaan 171, B-2020 Antwerpen, Belgium}%
\affiliation{Universidade Federal do Rio de Janeiro, Caixa Postal:68.528, Rio de Janeiro - RJ, 21941-972, Brazil}%
\author{F. M. Peeters}%
\email{francois.peeters@ua.ac.be}\homepage{http://www.cmt.ua.ac.be}%
\affiliation{Universiteit Antwerpen, Groenenborgerlaan 171, B-2020 Antwerpen, Belgium}%

\begin{abstract}
A thin superconducting disk, with radius $R=4\xi$ and height $H=\xi$, is studied in the presence of an applied
magnetic field parallel to its major axis. We study how the boundaries influence the decay of the order
parameter near the edges for three-dimensional vortex states.

\end{abstract}

\keywords{Vortex state,Mesoscopic sperconductor,Ginzburg-Landau}
\pacs{74.20.-z,74.60.-w,74.50.+r}
\maketitle
A few years ago the response of a mesoscopic superconducting disk to a magnetic field parallel to its major axis
was measured \cite{geim} and more recently the detection of giant vortices was made thanks to new advances in
small-tunnel-junction technology \cite{kanda}. The small volume to area  ratio of mesoscopic systems brings new
and interesting physical properties such as the onset of spontaneous and persistent currents \cite{kulik} and
for this reason the treatment of boundaries must be carefully considered.
\begin{figure}[h]
\centering
\includegraphics[width=\linewidth]{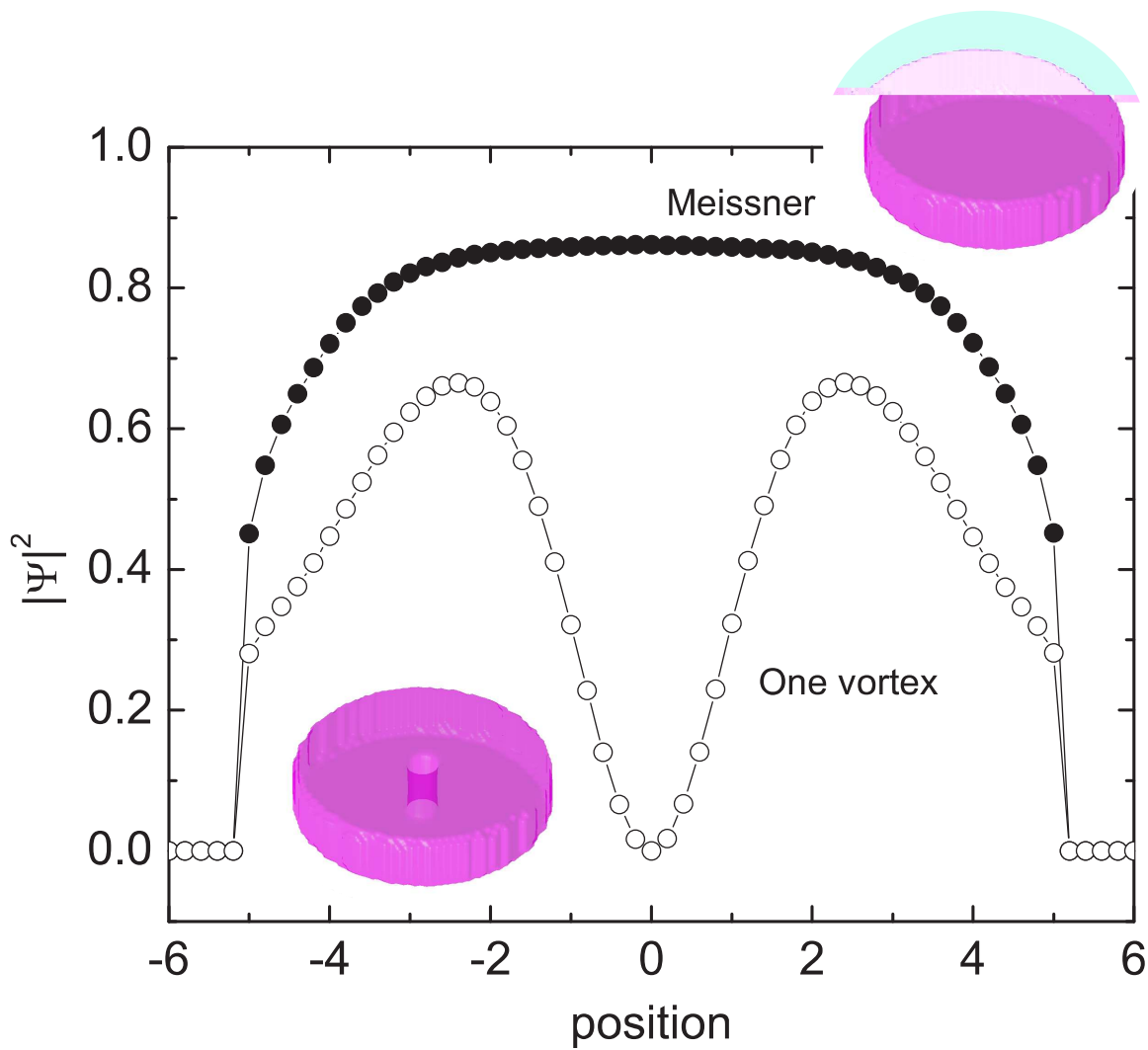}
\caption{$|\Psi|^2$ vs. the distance from the center of the disk is shown here for the case of zero and one
vorticity. The dots correspond to $|\Psi|^2$ values at mesh grid points. The three-dimensional figures
correspond to isocontours taken at 20$\%$ of its maximum value.}
\label{fig1}%
\end{figure}
In this work the three-dimensional Ginzburg-Landau (GL) theory is
solved in a parallelepiped cell that contains a smaller mesoscopic
superconducting disk inside. Space is discretized and gauge
invariance kept on a 61 by 61 by 16 grid. The distance between two
consecutive mesh points along any of the major axes is $\xi/5$,
$\xi$ being the coherence length at some fixed temperature. Thus the
disk occupies the center of a $12.0\xi$ by $12.0\xi$ by $3.0\xi$
cell. The vortex state solutions are obtained by numerical
minimization of the GL free energy in the cell through the method of
simulated annealing. Coupling of the disk to the outside
non-superconducting space is included and the order parameter
converges to zero outside the disk at the end of the minimization
procedure. This makes the present method somewhat different to those
that just seek the solution inside, but not outside, the disk
\cite{maurer}.
\begin{figure}[t]
\centering
\includegraphics[width=\linewidth, height=0.8\linewidth]{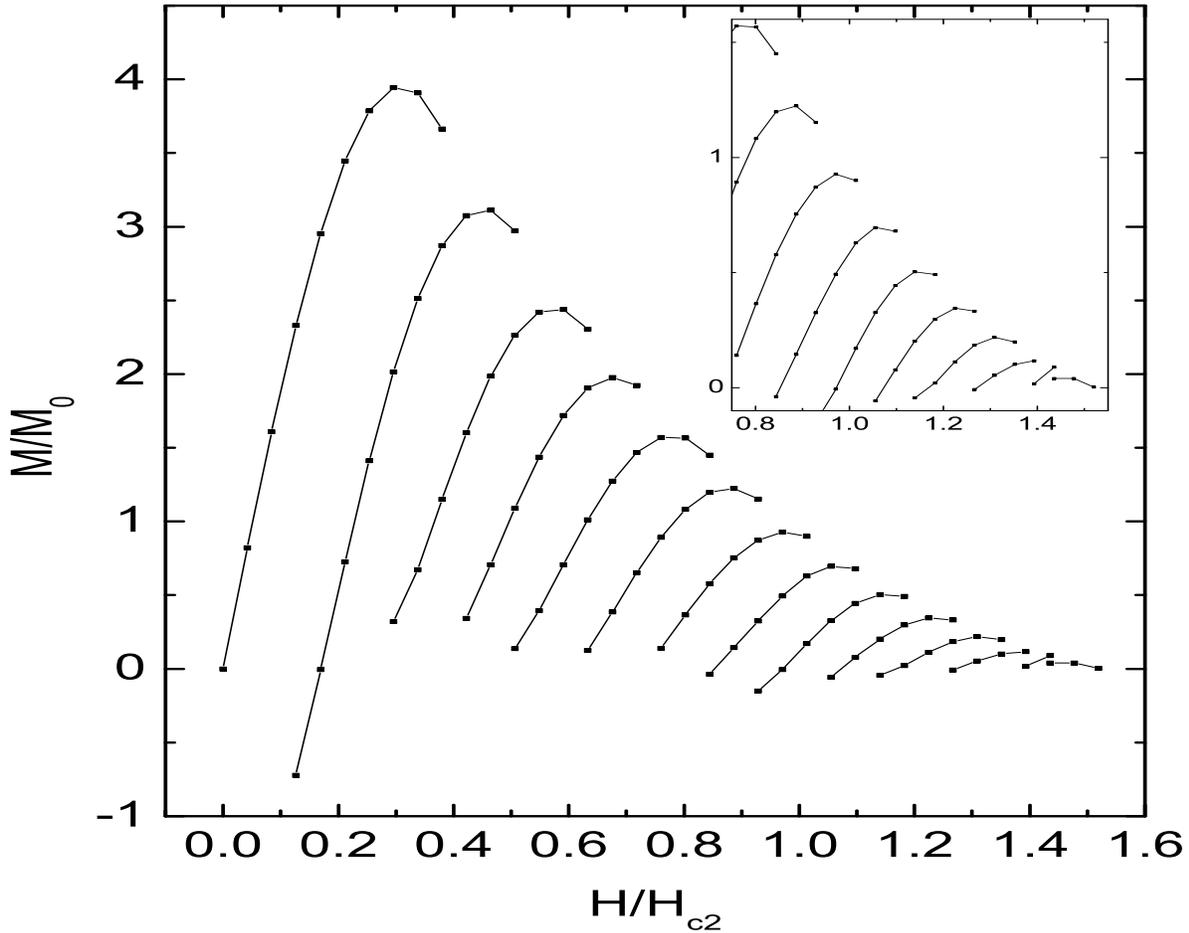}
\caption{Magnetization (arbitrary units) vs. the applied field. Inset shows the $H_{c2}$ to $H_{c3}$ region. The
magnetization is normalized by a {\it negative} arbitrary constant. }
\label{fig2}%
\end{figure}
\begin{figure}[t]
\centering
\includegraphics[width=\linewidth, height=0.8\linewidth]{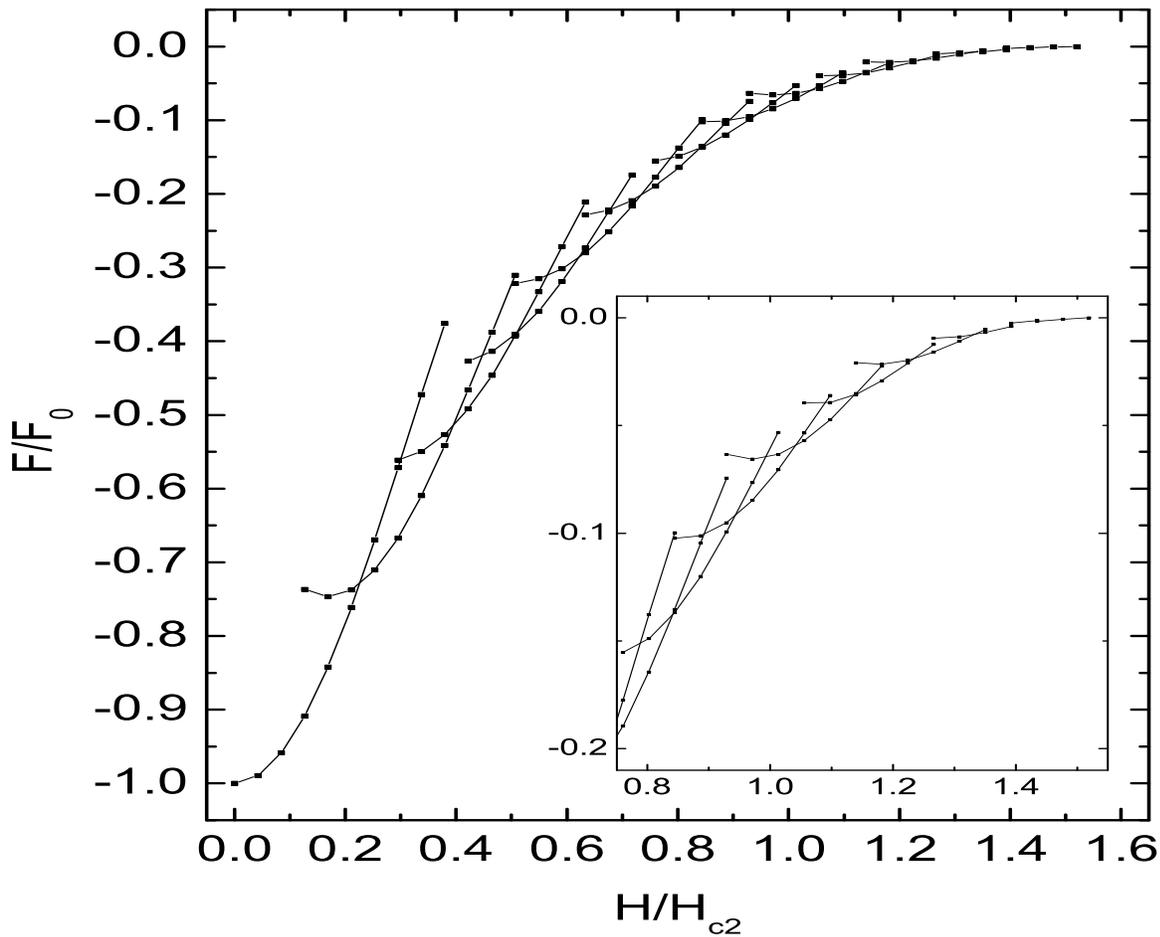}
\caption{Free energy vs. the applied field is shown here, the inset shows the $H_{c2}$ to $H_{c3}$ region. The
free energy is normalized to $H_c^2/8\pi$.}
\label{fig3}%
\end{figure}
The present description is restricted to a hard type II
superconductor whose free energy is simply given by, $f = \int
{{dv}\over{V}} \; (\tau \xi^2  |(\nabla-2\pi i/\phi_0
\textbf{A})\psi|^2 -\tau |\psi|^2 + {1 \over 2} |\psi|^4)$,
expressed in reduced units such that the density is normalized to
one. The magnetization is determined from ${\bf M} = const \int dv
\; {\bf x} \times {\bf J}$, where ${\bf J}$ is the supercurrent. The
shape of the mesoscopic superconductor enters directly into the free
energy through the step-like function $\tau(\bf{x})$, equal to one
inside the sample, and zero outside. As shown in Ref. \cite{doria}
this approach yields the de Gennes boundary condition. Here we make
the $\tau(\bf{x})$ function smooth with an exponential decay linking
the two sides. The condition of a vanishing supercurrent component
pointing outside the superconductor is only approximately enforced
between the disk and the outside external space: $\tau= \exp(-|({\bf
x-R})/\eta|^N)$. We take here for the two adjustable parameters,
$\eta=0.8\xi$ and $N=8$. Numerically, this smoothness is
advantageous since a steep $\tau(\bf{x})$ function directly couples
a mesh point in the disk to an outside point rendering a severe
depletion of the order parameter inside. In the present approach the
order parameter never reaches its maximum value of one, predicted
for a bulk superconductor, regardless of the $\eta$ and $N$ values
because the coupling to the outside conspires to lower the
superconducting density near the edge of the disk, as exemplified in
figure \ref{fig1} for the cases of none and one vortex inside the
disk. The magnetization and free energy curves obtained with the
present approach display a total of 13 lines as shown in figures
\ref{fig2} and \ref{fig3}, respectively. Each line corresponds to a
distinct vorticity state. The crossing of these lines in case of the
free energy defines the so-called matching fields \cite{schwei}. The
present results are in fair agreement to those found by Baelus
\cite{ben} for an extremely thin disk using a two-dimensional
approach for the superconducting density.

We have theoretically studied the vortex states of a mesoscopic disk with finite thickness. Previous studies
have also considered a mesoscopic disk but with a vanishing thickness\cite{ben}. We conclude that the present
minimization procedure of the Ginzburg-Landau free energy is able to obtain the vortex patterns of truly
three-dimensional mesoscopic superconductors.

\textbf{Acknowledgments} A. R. de C. Romaguera and M. M. Doria thank CNPq (Brazil), FAPERJ (Brazil) and the
Instituto do Mil\^enio de Nanotecnologia (Brazil) for financial support. F. M. Peeters acknowledges support from
the Flemish Science Foundation (FWO-Vl), the Belgian Science Policy (IUAP), the JSPS/ESF-NES program and the
ESF-AQDJJ network.


\begin{thebibliography}{7}
\bibitem{geim} A. K. Geim et al., Nature 390 (1997) 259.
\bibitem{kanda} A. Kanda et al., Phys. Rev. Lett. 93 (2004)
257002.
\bibitem{kulik} I.O. Kulik, L. Temp. Phys. 30 (2004) 531.
\bibitem{doria} M.M. Doria M.M \and Zebende G.F., Phys. Rev. B 66 (2002) 64519; M.M. Doria and Antonio R. de C. Romaguera,  Euro. Phys. Lett. 67 (2004) 446.
\bibitem{maurer} S.M. Maurer et al. Phys. Rev. B 54 (1996) 15372.
\bibitem{schwei} V. A. Schweigert et al. Phys. \ Rev. \ Lett. 81 (1998) 2783.
\bibitem{ben} B. J. Baelus and F. M. Peeters, Phys. Rev. B 65 (2002) 104515.
\end{thebibliography}
\end{document}